\documentstyle[preprint,aps]{revtex}
\begin{document}

\draft

\title{Current--Voltage Characteristics of Two--Dimensional
Vortex Glass Models}

\author{R.~.A.~Hyman{$^1$}, M.~Wallin{$^2$}, M.~P.~A.~Fisher{$^3$},
 S.~M.~Girvin{$^1$}, A.~P.~Young{$^4$}}
\address{
{$^1$}Department~of~Physics,
Indiana~University, Bloomington, IN~47405}
\address{
{$^2$}Department~of~Theoretical~Physics,
Royal~Institute~of~Technology, S-100~44~Stockholm, Sweden}
\address{
{$^3$}Institute for Theoretical Physics, University of California,
Santa Barbara, CA 93106}
\address{
{$^4$}Department of Physics, University of California, Santa Cruz, CA 95064}


\maketitle

\begin{abstract}
We have performed Monte Carlo simulations to determine
current--voltage characteristics of two different vortex glass models in two
dimensions. The results confirm the conclusions of earlier studies
that there is a transition at $T=0$. In addition we find that,
as $T\to 0$, the linear resistance vanishes
exponentially, and the current
scale,
$J_{nl}$, where non-linearities appear in the $I$--$V$ characteristics
varies roughly as $T^3$, quite different from the predictions of
conventional flux creep theory, $J_{nl} \sim T$.
The results for the two models agree quite well with each other,
and also agree fairly well
with recent experiments on very thin films of YBCO.
\end{abstract}

\section{Introduction}
Fluctuation effects are much larger in high $T_c$ superconductors than in
conventional materials~\cite{ffh}, so there has been considerable effort,
recently,
to go beyond the traditional mean field approximation in describing the
superconducting phase transition.
The effects of fluctuations, and also of disorder,
are particularly strong in a magnetic field.
In fact much of the $H-T$ phase diagram of high $T_c$ materials is
occupied by a ``vortex liquid'' regime in which the resistance has
dropped because superconducting short range order has formed, so flux lines
exist locally, but the resistance is not yet zero because the flux
lines move under the action of a
Lorentz force due to the current, and hence give rise to a
voltage~\cite{ks}. An important
question is whether, at lower $T$,
the flux lines can be collectively pinned by
defects sufficiently strongly
that they have no linear response to the Lorentz
force, which implies a vanishing linear resistance.
In the presence of disorder, there is
no long range order in the arrangement of vortices~\cite{lo}.
Nonetheless, it is argued~\cite{mpaf} that a transition
to a state with vanishing linear resistance can occur.
Such a state is called the vortex glass.

While there is theoretical~\cite{rtyf,other_dwrg}
and experimental~\cite{expts} evidence
for a finite vortex glass transition temperature, $T_c$, in bulk
superconductors, several simulations~\cite{fty,other_dwrg}
have clearly shown that $T_c=0$ in
two--dimensional systems, and rigorous analytic
arguments~\cite{nishimori} have established that there is no vortex
glass order at finite $T$ in 2-$d$.

Recent work\cite{bokil} has shown that
inclusion of gauge field fluctuations (ie, screening)
changes the universality class,
but one still has $T_c=0$ in two dimensions.
Even though $T_c = 0$, the correlation length
diverges as $T \to 0$,
which leads to observable consequences at finite temperatures, as we
shall discuss in detail here.
In contrast,
a  different simulation of the non-linear IV characteristic of the d=2
gauge glass model found evidence for a finite $T_c$.\cite{YHLi}.
However behavior consistent with $T_c = 0$  has
recently been observed~\cite{expt_2d} in experiments on very
thin, $16 \AA$, films of YBCO.

Although the experiments~\cite{expt_2d} on the 2-$d$ samples
are in quite good agreement
with theory, they measure different quantities from what has
been calculated so far, and a scaling hypothesis is needed to make a
comparison. The experiments determined current--voltage (I--V)
characteristics while the simulations investigated the size--dependence
of the rigidity of the system with respect to a twist in the phase of
the condensate. It therefore seems worthwhile to make a
{\em direct} comparison by calculating I--V characteristics from
the simulations on 2-$d$ systems. As an additional benefit, we study {\em two}
models which are somewhat different microscopically: the gauge glass,
defined in Eq.~(\ref{ham:gg}) below, which has effectively {\em random} fluxes
penetrating the sheet, and a more realistic model, defined in
Eq.~(\ref{ham:rand_pot}) below, which
has a net {\em uniform} field penetrating the sheet and a
random pinning potential
for the vortices. We find that universal properties are the same
for these two models. Since the gauge glass has been extensively used
for numerical studies of the vortex glass transition it is reassuring
that it gives the same results as a model with a net field,
at least in two--dimensions.

\section{The Models}
The first model that we study is the gauge glass~\cite{rtyf,other_dwrg,fty},
whose Hamiltonian is
\begin{equation}
{\cal H}_{\rm gg} = - \sum_{\langle i,j\rangle}
\cos ( \phi_i - \phi_j - A_{ij}) \quad .
\label{ham:gg}
\end{equation}
The phase of the condensate,
$\phi_i$, is defined on each site, $i$, of a square lattice, with
$N = L^2$ sites.
The sum is over all nearest neighbor pairs on the
lattice. The effects of the external magnetic field and disorder are
represented by the quenched vector potentials, $A_{ij}$, which are
taken to be independent random variables with a uniform distribution
between 0 and $2\pi$.

To compute the I--V characteristics we need to incorporate dynamics into
the model. The standard way of doing this is to
view the model as a set of
coupled Josephson junctions~\cite{ls,shenoy,monteitel}.
Josephson's and Kirchoff's equations for the current are then
\begin{eqnarray}
\label{current}
I_{ij} & = & {V_{ij} \over R_0} + I_c \sin( \phi_i - \phi_j - A_{ij})
+ \eta_{ij}(t) \\
V_{ij} & = & {\hbar \over 2 e} {d \over d t}(\phi_i - \phi_j)
\label{josephson}
\\
\sum_j I_{ij} & = & I_{i:{\rm ext}} \quad ,
\label{kirchoff}
\end{eqnarray}
where $i$ and $j$ are nearest neighbor pairs.
Eq. (\ref{current}) expresses the sum of the current from site $i$ to
neighboring site $j$ as the sum of a resistive current given by $V_{ij} /
R_0$, a Josephson current, and a Langevin current noise source
$\eta_{ij}(t)$. $I_c$ is the maximum Josephson
current of the nearest neighbor pair, and $R_0$ is the
shunt resistance of the pair.
The thermal noise has a Gaussian distribution with the following
properties:
\begin{eqnarray}
\langle \eta_{ij}(t) \rangle & = & 0 \quad , \\
\langle \eta_{ij}(t) \eta_{kl}(t^\prime) \rangle & = & {2 k_B T \over
R_0} \delta_{ij,kl} \delta ( t - t^\prime) \quad ,
\end{eqnarray}
which ensures that the system comes to thermal equilibrium at temperature
$T$.
Eq. (\ref{josephson}) is the Josephson relation connecting the
voltage $V_{ij}$ with the time derivative of the phase difference,
$\phi_i -\phi_j$. Eq. (\ref{kirchoff}) is Kirchoff's law expressing current
conservation at site $i$. $I_{i:{\rm ext}}$ is the external current at site
$i$. This is zero except for sites on the top row where
an external current $J \equiv I / L$,
is fed in, and sites on the bottom row where
the same current is extracted. The {\em total} current through the sample is
then $I$. The
average voltage across the system, $V$, is given by
\begin{equation}
V ={1\over L^2}{\hbar \over 2 e}
\sum_{i\in \mbox{\footnotesize bottom}}
\sum_{j\in \mbox{\footnotesize top}}
{d \over d t}\langle \phi_i - \phi_j \rangle
\end{equation}
where the brackets, $\langle \ldots \rangle$, denote a time average.
The average electric field is then obviously given by $E = V / L$.
For this model
we work in units where $\hbar / (2e) = R_0 = I_c = 1$.
Throughout the paper we also set Boltzmann's constant to be unity.
The equations of motion are integrated using a first order approximation
with a time step of $\delta\tau = 0.05\tau$, where
$\tau = \hbar/(2eR_0I_c)$
is the basic unit of time (which is set equal to unity).
The scale of $\tau$ is the typical time for a
neighboring pair of sites to accumulate a relative phase of order unity.

We have also studied an equivalent form for the gauge glass written in terms
of vortices. To obtain this we
replace the cosine by the periodic Gaussian (Villain)
function~\cite{villain}, and perform
standard manipulations~\cite{villain,vortex_model}, obtaining
\begin{equation}
\label{ham:gg:vortex}
{\cal H}_{\rm gg}^V = -{1\over 2} \sum_{i,j}
(n_i - b_i)G(i - j) (n_j - b_j) \quad ,
\end{equation}
where the $\{n_i\}$ run over all integer values, subject to the
``charge neutrality'' constraint $\sum_i n_i = 0$, and are interpreted
as the vortex ``charges'', and $G(i - j)$ is the vortex interaction,
\begin{equation}
\label{vortex_int_2d}
G(i - j) =
\left( {2 \pi \over L }\right)^2 \sum_{{\bf k} \ne 0}
{1 - \exp[i {\bf k} \cdot ({\bf r_i} - {\bf r_j})] \over
4 - 2 \cos k_x - 2 \cos k_y} \quad.
\end{equation}
At large distance, $G(i-j) \to 2 \pi \log |{\bf r_i} - {\bf r_j}|$.
The vortices sit on the sites of the dual lattice, which lie
in the centers of the squares of the original lattice.
The magnetic fluxes $b_i$ are the lattice curl of the vector potential and
are given by (1/$2 \pi$) times
the directed sum of the quenched vector potentials on
the links of the original lattice which surround the site $i$ of the
dual lattice.

We study the I--V characteristics of this vortex model by using Monte
Carlo dynamics.  That is, we equate Monte Carlo `time' and real time, an
approximation which is expected to be good in the limit of overdamped
dynamics and which has proven reasonable in other
simulations.\cite{gw}
Choosing a nearest neighbor pair, $i,j$ at random, we
try to increase $n_i$ by 1 and decrease $n_j$ by 1, thus transferring a
unit vortex from $j$ to $i$. If the change in energy is $\Delta E$, the
move is accepted with the probability
$1/(1 + \exp(\beta \Delta E))$ appropriate to the ``heatbath'' algorithm.
An applied
current density $J$ gives a Lorentz force of $J h / 2e$ on a unit
vortex. This can be incorporated into the Monte Carlo moves~\cite{gw}
by adding to $\Delta E$ an amount $Jh / (2e)$ if the vortex moves in
the opposite direction to the Lorentz force, subtracting this amount
if it moves
in the same direction, and making no change in $\Delta E$ if it moves in a
perpendicular direction. Biasing the moves in this way
takes the system out of equilibrium and causes a net flux
of vortices in a direction perpendicular to the current. This then
generates a voltage $V$, where
\begin{equation}
V = {h \over 2 e} \langle I^V (t) \rangle \quad ,
\end{equation}
with
\begin{equation}
I^V (t) = {1\over L \Delta t} \sum_i \Delta Q_i^V (t)
\end{equation}
the vortex current.
Here $t$ denotes a Monte Carlo ``time'' (incremented by $\Delta t$
after each attempted move), and $\Delta Q_i^V(t) = 1$ if a vortex at site $i$
moves one lattice spacing in the direction of the Lorentz force at time
$t$, $\Delta Q_i^V(t) = -1$ if the vortex moves in the direction opposite to
the Lorentz force, and $\Delta Q_i^V(t) = 0$ otherwise.
We set $\Delta t = 1/ 4N$ so that an
attempt is made to move each vortex once in each direction, on average,
per unit time.
We shall use units where $h / 2e = 1$ when dealing with vortex models.

The linear resistance can also be obtained from the Kubo formula for
fluctuations in the voltage in the absence of any net current. This is
exact for discrete time Monte Carlo dynamics provided the sum over time
is made symmetrical about $t=0$~\cite{y_orbach} i.e.
\begin{equation}
R_{\rm lin} = {1\over 2 T} \sum_{t=-\infty}^\infty \Delta t\ \langle
V(t) V(0) \rangle \quad ,
\end{equation}
which, in our units, can be expressed in terms of the vortex current as
\begin{equation}
\label{kubo}
R_{\rm lin} = {1\over 2 T} \sum_{t=-\infty}^\infty \Delta t\ \langle
I^V(t) I^V(0) \rangle \quad .
\end{equation}

Using Monte Carlo dynamics should be a good approximation near
a critical point, where the vortex motion is slow and overdamped.
However, because of discretization of time, and the fact that the
fastest that a vortex can move is one lattice spacing per time step, it is
not very satisfactory for large
currents or high temperatures, where the vortex motion is ballistic.
For example, at high temperatures with no bias current, a vortex
moves in the $\pm$ x direction with probability, 1/4, so, from Eq.
(\ref{kubo}), the resistance is given by
$R_{\rm lin} = 1/ (2 T)$ and
tends to zero, which is unphysical.

The gauge glass represents a system with {\em random} fluxes penetrating the
film but with zero net field. It is a convenient model to study but it
should be verified that it is in the same universality class as
experimental systems which have a {\em net} uniform
field penetrating the film and a random pinning potential for the vortices. We
have therefore also studied
a random pinning model with the following Hamiltonian,
\begin{equation}
{\cal H}_{rp} = -{1\over 2}\sum_{i,j} n_i G(i-j) n_j - \sum_i v(i) n(i)^2
\quad ,
\label{ham:rand_pot}
\end{equation}
where the $n(i)$ are restricted to the values $0, \pm 1$,
$G(i-j)$ is given by Eq.~(\ref{vortex_int_2d}), and $v(i)$ is
a random pinning potential, uniformly
distributed in the interval $-\Delta < v(i) < \Delta$. We set $\Delta =
\pi$ and fix the
net filling, $f \ ( \equiv (1/N) \sum_i n_i \ )$,
which effectively determines the
magnetic field, to be 1/4. We obtain I--V characteristics from
Monte Carlo dynamics.
At each Monte Carlo time we try to insert a $+1, -1$
pair at a randomly chosen pair of sites $i$ and $j$. The analysis is
then precisely the same as described above for the vortex representation of the
gauge glass.

\section{Scaling Theory}
To analyze the results it is necessary to understand how the I--V
characteristics vary in the vicinity of a second order
phase transition. A detailed scaling theory
has been developed~\cite{ffh} and we now summarize the results of this
for the case of a zero temperature transition, where the correlation
length diverges as
\begin{equation}
\label{xi}
\xi \sim T^{-\nu} \quad,
\end{equation}
and the relaxation time, $\tau$, also diverges. Normally one defines a
dynamic exponent, $z$, by $\tau \sim \xi^z$, but since the transition
is at $T=0$, the relaxation has an activated form and diverges
exponentially as $T \to 0$. Formally this corresponds to $z = \infty$.

The vector potential, ${\bf A}$, enters the Hamiltonian in Eq.(\ref{ham:gg})
in the dimensionless form $A_{ij} \sim \int_j^i {\bf A(r)}\cdot d{\bf r}$.
Therefore ${\bf A}$ scales as $1/\xi$.
The electric field is given by ${\bf E} = -\partial_t {\bf A}$
and so scales as
$1/(\xi \tau)$. Now $J E$ is the energy dissipated per unit volume per
unit time. The natural unit of energy is $k_B T$ so ${\bf E\cdot J}$
scales like $k_B T / (\xi^d \tau)$ and hence $J$ scales like
$k_B T/ \xi^{d-1}$. It is important to keep the factors of $T$ because
$T_c = 0$. Combining these results we obtain (for $k_B = 1$)
\begin{equation}
T {E \over J } {\tau \over \xi^{d-2}}
= g\left({ J \xi^{d-1} \over T } \right) \quad ,
\end{equation}
where $g$ is a scaling function.
In $d=2$ with $T_c = 0$ this becomes
\begin{equation}
T \tau {E \over J }
= g\left({ J \over T^{1 + \nu} } \right) \quad .
\label{EJscale}
\end{equation}
 From Eq.~(\ref{EJscale}) one sees that the characteristic current scale,
$J_{nl}$, at which non-linear behavior sets in, varies with $T$ as
\begin{equation}
\label{jnl}
J_{nl} \sim T^{1 + \nu} \quad .
\end{equation}

Now the linear resistance is just
\begin{equation}
R_{\rm lin} = \lim_{J \to 0} {E \over J} \quad,
\end{equation}
and $g(0)$ must be a constant, which we take to be unity, so we can write
\begin{equation}
\label{E-J}
{E \over J R_{\rm lin} }
= g\left({ J \over T^{\nu + 1} } \right) \quad ,
\end{equation}
and
\begin{equation}
\label{rlin}
T R_{\rm lin} = {1\over \tau} = A \exp\left(- \Delta E(T) / T \right) \quad,
\end{equation}
where $\Delta E(T)$ is the typical barrier that a vortex has to cross to
move a distance $\xi$. One conventionally defines a barrier exponent
$\psi$ by $\Delta E \sim \xi^\psi \sim T^{-\psi\nu}$ in terms of which
\begin{equation}
\label{rlin_psi}
T R_{\rm lin} \sim \exp\left(- C / T^{1 + \psi \nu}\right) \quad.
\end{equation}
It has been suggested~\cite{ffh} that $\psi$ may be zero in $d = 2$,
leading to barriers which are either finite as $T \to 0$ or diverge
logarithmically. In the latter case the linear resistance would vary as
$\exp\left( - C |\log(T)|^{\mu}/ T\right)$, where $\mu$ is another
exponent.

In Eq.~(\ref{rlin}), the linear resistance is seen to vanish
exponentially fast as $T \to 0$. In these circumstances it is generally
more difficult to estimate the form of possible power law prefactors than the
form of the leading exponential dependence. We have incorporated a
factor of $T$ on the LHS of Eq.~(\ref{rlin}) because it emerged
naturally from the scaling ansatz. This factor of $T$ also looks
reasonable when compared with the Kubo formula, Eq.~(\ref{kubo}), since
it is $T R_{\rm lin}$ which is given by the voltage fluctuations.
However, it is possible that additional factors of $T$ could be present
in the scaling region.

In a finite system, the I--V characteristics will also depend on the
size of the system when the bulk correlation length,
$\xi$, becomes comparable with $L$. According to
finite size scaling~\cite{priv} only the ratio $L/\xi$ is important and
so, from Eq.~(\ref{xi}), one can generalize
Eq.~(\ref{E-J}) to
\begin{equation}
\label{E-J-L}
{E \over J R_{\rm lin} }
= \tilde{g}\left({ J \over T^{1 + \nu} },\ L^{1/\nu} T \right) \quad .
\end{equation}
%
The non-linear behavior in the finite-size regime, described by
Eq.~(\ref{E-J-L}), is rather complicated because it involves a function
of two variables. We have therefore studied non-linear behavior either in
the range where finite size corrections are negligible or, having already
determined $\nu$, choose $L$ and $T$ such that the second argument,
$L^{1/\nu} T$, is constant.

\section{Results}
In Fig.~\ref{expt_res} we show some of the experimental data of Dekker et
al.~\cite{expt_2d} on $16\AA$ films of YBCO. For small current
densities, the data is flat, indicating Ohm's law, with a linear
resistivity which decreases rapidly with decreasing
temperature. As $J$ is
increased the data starts to curve upwards, indicating non-linear
response. The current scale, $J_{nl}$,
at which non-linear behavior sets in, is also seen to decrease with
decreasing temperature. Analyzing all their data, Dekker et al. find the linear
resistivity varies as
\begin{equation}
R_{\rm lin} \sim \exp ( - C/ T^p)
\end{equation}
with $p \simeq 0.6$. This variation, which
is less rapid than an Arrhenius form, $p=1$, is difficult to understand
from the classical models which we study here. It may therefore indicate
that quantum tunneling of vortices is important at the
lowest temperatures. On the other hand,
$J_{nl}$ is found to vary with $T$ like $T^3$,
which, from Eq.~(\ref{jnl}) implies $ \nu \simeq 2$, in agreement with
the earlier simulations~\cite{fty,other_dwrg} which studied
a classical model.

We next discuss our numerical results for the I-V characteristics
of the gauge glass in the vortex representation. Data for
$T$ times the linear
resistance against $1/T$ is shown in Fig.~\ref{gg_v_rlin}
on a log-linear plot. A
simple Arrhenius form, i.e. a temperature independent barrier height,
$\Delta E$, would correspond to a straight line. In fact for the largest
size, the data is very close to a straight line indicating that the
barrier exponent, $\psi$,
in Eq.~(\ref{rlin_psi}) is zero or close to it.
These results are consistent the suggestion~\cite{ffh} of a
logarithmically increasing barrier,
but this weak dependence will be
difficult to see on data which are
over a modest range of
temperature 
in finite size systems,    

Data for the non-linear response at finite $J$ is shown in
Fig.~\ref{gg_v_E-J}. The data for the smallest current density,
$J$, was actually obtained for $J=0$ from the Kubo
formula, Eq.~(\ref{kubo}). As in the experimental results in
Fig.~\ref{expt_res}, one sees Ohm's law at small
$J$ (where each dataset is horizontal) with a linear resistance
which decreases rapidly with
temperature, as shown in more detail
in Fig.~\ref{gg_v_rlin}, but deviations from Ohms law occur
at a scale, $J_{nl}$,
where the data start to curve upwards. One sees that
$J_{nl}$ decreases with decreasing temperature, as expected.
%
 From Fig.~\ref{gg_v_rlin}, finite size
effects appear quite small for
$L=16$ at $T= 0.5$ and 0.8. Assuming $\nu \approx 2$, the data for
$L=32 $ in Fig.~\ref{gg_v_rlin}
at $T= 0.35$ should also not be significantly affected by finite size
effects. We have therefore analyzed the data in Fig.~\ref{gg_v_E-J}
according to the expected result for bulk behavior, Eq.~(\ref{E-J}). The
scaling plot is shown in Fig.~\ref{gg_v_E-J_scale}. The data scales
reasonably well with $\nu = 1.8$ which is in quite good agreement with
other estimates~\cite{fty,other_dwrg}.
The values of $R_{\rm lin}$ in this plot were obtained
from the Kubo formula.

Next we discuss the data obtained for the phase representation of the
gauge glass.  Results for the linear resistance are shown in
Fig.~\ref{gg_rlin}. The data is consistent with an Arrhenius form for the
largest sizes, as was found for the vortex representation
(Fig.~\ref{gg_v_rlin}).
%
%
Results for non-linear current voltage characteristics for the gauge
glass model in the phase representation are shown in Fig.~\ref{gg_E-J}.
Unlike the vortex representation, which has discrete time and so a
maximum vortex velocity, the dynamics
of the phase representation uses continuous time\cite{rosscomment}
and so can sensibly
be applied for large values of $J$ (and also high-$T$). In
Fig.~\ref{gg_E-J} one clearly sees a ``flux-flow'' regime for large $J$,
where the Lorentz force is sufficient to overcome pinning, and the only
hindrance to vortex motion comes from friction. This leads to a
resistance which is roughly independent of $J$ and also only rather weakly
dependent on temperature. For small $J$, however, the vortices are
pinned by defects and move only by activation over barriers.
The linear resistance in this ``flux-creep' region, is therefore
much smaller than that observed for larger $J$ and is also strongly
temperature dependent.


We next describe our results for the random pinning potential model
in Eq.~(\ref{ham:rand_pot}). The linear resistance is shown in
Fig.~\ref{rp_rlin}. As for the gauge glass, the data for the
larger sizes seem to be
tending towards an Arrhenius behavior.


The non-linear behavior of the random pinning potential model is shown in
Fig.~\ref{rp_E-J}. As for the  gauge glass results in
Fig.~\ref{gg_E-J}
one sees a ``flux-flow'' regime
for large $J$, and a ``flux-creep'' region at small $J$ where the linear
resistance is small and strongly temperature dependent. Deviations
from the Ohm's law behavior in the flux creep regime
occur at a current scale, $J_{nl}$, which decreases as $T$ decreases.
For sufficiently large $J$, $E$ will saturate, because there is a
maximum
vortex velocity when each vortex hops every step, and so the ratio $E/J$
will decrease.
 This (unphysical) behavior, which is just visible in the figure
at the largest values of $J$, is
caused by discretization of time
in the Monte Carlo simulations. This discretization is, however, {\em not}
expected to affect universal
critical properties near the $T=0$ critical point.

A scaling plot of the non-linear I-V characteristics of the random
pinning potential model is shown in Fig.~\ref{rp_E-J_scale}.
Since there are finite-size effects within the range of accessible
sizes, we assume that
$\nu \simeq 2$, and choose sizes and temperatures such that
$L T^2$ is constant. In this way, the second argument in the scaling
function in Eq.~(\ref{E-J-L}) is constant and $E/(J R_{\rm lin})$
should only be a
function of $J/T^3$.
The data is seen to scale very well over a wide
range. These results provide strong evidence that the gauge glass and random
pinning potential models are in the same universality class with a correlation
length exponent at the $T=0$ transition of $\nu \simeq 2$.

Finally, in Fig.~\ref{expt_theory},
we compare the non-linear current-voltage characteristics from the
simulations on the random pinning potential and gauge glass models
with the experimental results shown in Fig.~\ref{expt_res}. The
same value of $\nu=2$ was used for all the data, and, for each system, a
temperature scale, $T_0$, was adjusted to get the best scaling. The two
sets of simulation data agree quite well with each other.
While the gauge glass data were obtained in the region where $\xi \ll L$, as
deduced from the results for $R_{lin}$, the random pinning potential model data
was in the finite size region and so the data were taken
at fixed $L T^2`$ (or equivalently at fixed, but not very small, $\xi/ L$).
The good agreement between the two sets of data
indicates that finite size corrections are not very important for the
I-V characteristics. There is, however, a discrepancy between the numerical
results and experiment, for which we do not have an explanation.

\section{Conclusions}
We have studied the I-V characteristics of two models for vortex glass
behavior in two dimensions. For the first model, the gauge glass, our
results confirm earlier studies\cite{other_dwrg,fty} which found a zero
temperature transition with a correlation length exponent $\nu \simeq
2$.
This behavior has also been seen experimentally~\cite{expt_2d}
on very thin films of YBCO, though the detailed form of the I-V scaling
function is somewhat different between theory and experiment.
We also find that the linear resistance varies with an Arrhenius form as
$T \to 0$, indicating that the barrier exponent, $\psi$, is either zero or
close to zero. Experimentally the resistance appears to vanish at low
$T$ less rapidly than an Arrhenius form,
which may indicate that quantum fluctuations of the vortices play a role in the
YBCO films.
The other model which we study,
the random pinning potential model, is rather
more realistic in that it describes a system with a uniform applied
magnetic field perpendicular to the film, as opposed to the gauge glass
which has random fluxes. Nonetheless, the random pinning potential model is
found to
be in the same universality class as the gauge glass, since they both
have a
zero temperature transition with $\nu \simeq 2$ and an Arrhenius
behavior for the linear resistance.
Hence these two models are equivalent in $d=2$.
It can also be shown that
in $6-\epsilon$ dimensions,
the gauge glass transition is in the
same universality class as the vortex glass transition in a disordered
model with a net magnetic field.
It is therefore quite plausible that the two models have the same
critical behavior in three dimensions, though we are not aware of a
direct demonstration of this.

Acknowledgements: We should like to thank C.~Dekker for providing us
with the data in Fig.~\ref{expt_res}.
APY is supported by NSF grants DMR 91-11576
and DMR-94-11964.  MPAF is supported by NSF grants PHY89-04035
and DMR-9400142.  SMG and RAH are supported by DOE Grant DE-FG02-90ER45427.
MW is supported by the Swedish Natural Science Research Council.

\begin{figure}
\caption{
A plot of some of the experimental data, at a field of 0.5 T,
on very thin, $16 \AA$, films of
YBCO from Ref.~\protect\cite{expt_2d}.
The numbers denote the temperatures, in Kelvin, at which the
different data sets were taken.
}
\label{expt_res}
\end{figure}

\begin{figure}
\caption{
A plot of $T$ times the linear resistance, on a logarithmic scale, against
$1/T$
for different sizes for the gauge glass in the vortex representation.
The data for the largest size, $L=32$, is well approximated by a straight line,
indicating a temperature independent barrier height, i.e. Arrhenius
behavior.
}
\label{gg_v_rlin}
\end{figure}

in

\begin{figure}
\caption{
A log-log plot of the non-linear
current voltage characteristics of the gauge glass model in
the vortex representation. The data shown for the smallest current density,
$J$, is actually for $J=0$ and was obtained from the Kubo
formula, Eq.~(\protect\ref{kubo}). One sees Ohm's law at small
$J$ (where each dataset is horizontal) with a linear resistance
which decreases rapidly with
temperature, as shown in more detail in
Fig.~\protect\ref{gg_v_rlin}.
Deviations from Ohm's law, where the data
start to curve, occur
at a current scale, $J_{nl}$, which decreases with decreasing temperature.
}
\label{gg_v_E-J}
\end{figure}

\begin{figure}
\caption{
A scaling plot of the data in Fig.~\protect\ref{gg_v_E-J},
assuming that finite-size corrections are small and hence that the scaling form
expected for bulk behavior, Eq.~(\protect\ref{E-J}), is
appropriate. The value $\nu = 1.8$
obtained from this fit, is is reasonable agreement with other estimates.
}
\label{gg_v_E-J_scale}
\end{figure}

\begin{figure}
\caption{
A plot of the linear resistance, on a logarithmic scale, against $1/T$
for different sizes for the gauge glass in the phase representation.
The data for the largest size, $L=16$, is fairly close to a straight line,
indicating a temperature independent barrier height, i.e. Arrhenius
behavior. This is similar to what was observed in the vortex
representation, Fig.~\protect\ref{gg_v_rlin}.
}
\label{gg_rlin}
\end{figure}


\begin{figure}
\caption{
Results for non-linear current voltage characteristics for the gauge
glass model in the phase representation. One sees a ``flux-flow'' regime
for large $J$, where the resistance is largely independent of $J$ and
$T$, and a ``flux-creep'' region at small $J$ where the linear
resistance is small and strongly temperature dependent. Deviations
from the Ohm's law behavior seen at small $J$ occur at a current scale,
$J_{nl}$, which decreases as $T$ decreases.
}
\label{gg_E-J}
\end{figure}

\begin{figure}
\caption{
A plot of the linear resistance, on a logarithmic scale, against $1/T$
for different sizes for the random pinning potential model in
Eq.~(\protect\ref{ham:rand_pot}).
The data for the largest sizes seems to be tending to a
straight line,
indicating a temperature independent barrier height, i.e. Arrhenius
behavior. This is similar to what was observed for the gauge glass,
see Figs.~\protect\ref{gg_v_rlin} and \protect\ref{gg_rlin}.
}
\label{rp_rlin}
\end{figure}

\begin{figure}
\caption{
Results for non-linear current voltage characteristics for the
random pinning potential model in Eq.~(\protect\ref{ham:rand_pot}).
For each size, $L$, the temperature has been chosen so that
$L T^2 = 2$, in order to keep the second argument of the finite-size
scaling function in Eq.~(\protect\ref{E-J-L}) constant. The net
filling, $f \ ( \equiv (1/N) \sum_i n_i \ )$, is equal to 1/4.
As for the gauge glass results presented in Fig.~\protect\ref{gg_E-J}
one sees a ``flux-flow'' regime
for large $J$, and a ``flux-creep'' region at small $J$ where the linear
resistance is small and strongly temperature dependent. Deviations
from the Ohm's law behavior seen in the flux creep regime
occur at a current scale, $J_{nl}$, which decreases as $T$ decreases.
}
\label{rp_E-J}
\end{figure}

\begin{figure}
\caption{
A scaling plot of the non-linear current-voltage characteristics of the
random pinning potential model.
Since there are finite-size effects in this data, we assume that
$\nu \simeq 2$, and choose sizes and temperatures such that
$L T^2$ is constant. In this way, the second argument in the scaling
function in Eq.~(\protect\ref{E-J-L})
is constant and $E/(J R_{lin})$ should only be a
function of $J/T^3$.
The data is seen to scale very well to this expected form over a wide
range.
}
\label{rp_E-J_scale}
\end{figure}

\begin{figure}
\caption{
A scaling plot combining the non-linear current-voltage characteristics of
both the experimental data in Fig.~\protect\ref{expt_res}
(lines) and simulations (points). The data
points with $L T^2 = 2$
were obtained for the random pinning potential model and the other two
sets of data points, at $T = 0.35$ and $0.50$, were for obtained
for the gauge glass model in the
vortex representation. The value $\nu = 2$ was used for all the data.
The temperature scale was set by $T_0 = 1$ (random pinning potential model),
$T_0 = 1.15$ (gauge glass), and $T_0 = 12 K$ (experiment).
The two sets of data from the simulations agree quite well with each other,
but the experimental results lie lower than theory
at large values of $J/T^{1+\nu}$.
}
\label{expt_theory}
\end{figure}


\begin{references}
\bibitem{rosscomment}  Of course our numerical implementation of the continuous
time dynamics uses a discrete time step, but this time step
is much smaller than
the characteristic microscopic time, $\tau$, which corresponds to the
time of one sweep in the Monte Carlo simulation.
\end{references}
\end{document}